\documentstyle[12pt]{article}
\def \as {\alpha_s}
\def \a  {\alpha}
\def \ao {\alpha_0}

\def \ms {{\overline{MS}}}

\newcommand{\prepr}[1] {\begin{flushright} {\bf #1} \end{flushright} \vskip
1.5cm}
\newcommand{\titul}[1] {\begin{center}{\large\bf #1 } \end{center}\vskip 1.cm}
\newcommand{\autor}[1] {\begin {center} {\large \lineskip .5em #1 }
                        \end   {center} }
\newcommand{\lugar}[1] {\begin{center} {\it #1} \end{center}}
\newcommand{\abstr}[1] {{\begin{center} \vskip .5cm {\bf Abstract
                        \vspace{0pt}} \end{center}}\begin{quote} #1
                        \end{quote}}

\begin{document}

\begin{titlepage}
\prepr{UCI-TR/94-4\\ May 1994}
\titul{Next-to-next-to-leading order QCD analysis of DIS
structure functions}
\autor{G. Parente}
\lugar{Department of Physics, University of California  \\
Irvine, CA 92717, USA}
\autor{A.V. Kotikov and V.G. Krivokhizhin}
\lugar{Laboratory of Particle Physics, Joint Institute for Nuclear Research\\
101 000 Dubna, Russia}
\abstr{ It is performed for the first time
a next-to-next-to-leading order
analysis of deep inelastic
structure functions $F_2$ and $F_L$
using the recently determined first moments
of the
non-singlet anomalous dimensions and
the corresponding Wilson
coefficients. From a comparison to BCDMS data we found a
slight decrease in the energy scale of perturbative
QCD with respect to the next-to-leading order analysis,
which brings the scale closer to the value predicted
from a scheme invariant approach.}
\end{titlepage}
\newpage



The determination of the QCD
coupling constant $\alpha_s$ at some energy is one of the present
problems of both experimental and theoretical high energy physics.
Deep inelastic lepton-nucleon scattering (DIS) has been
one of the most important processes for qualitative tests
of perturbative QCD and
in particular for the
determination of the energy scale $\Lambda$
involved in $\alpha_s$.
In the last few years
quite exact experimental data for DIS structure functions
has been obtained (see \cite{1}, for review) over a wide range of
the Bjorken variable $x$. Theoretical knowledge had been restricted
to next-to-leading order (NLO) approximation
(see \cite{2} and its references)
except sum rules \cite{3} and the lower (Mellin)
moment of the longitudinal
structure function
$F_L$
\cite{4} which have been known with next-to-next-to-leading (NNLO) accuracy.

The purpose of this letter is to present a NNLO analysis of structure
functions. The two new ingredients needed in the calculation,
i.e. the anomalous
dimensions at three loops and the
Wilson coefficients of the quark and gluon operators at
two and three loops
for the transverse and longitudinal parts, respectively,
were recently calculated by Larin,
van Ritbergen  and Vermaseren: The first
ten O($\as^2$) moments of the $F_2$ coefficients
\cite{5}\footnote{ The corresponding kernels
into Mellin convolutions with parton distributions
were calculated
by van Neerven and Zijlstra \cite{6}},
and the n = 2, 4, 6 and 8 moments for the non-singlet (NS)
anomalous dimension at NNLO,
and also the $F_L$ coefficients at O($\as^3$) \cite{7}.

Some attempts have also been
made in order to include the effect of higher
order corrections in structure functions using the information of the
NLO and the invariance on
the renormalization scheme of the physical results. Here we extend
one of these scheme invariant approaches with the inclusion
of NNLO terms.


At three loops the Mellin moments of structure functions
at two different
scales $Q^2$ and $Q_0^2$ are related by\footnote{
Through this work we use
the short notation $\a = \alpha_s(Q^2)/4\pi$ and
$\ao = \alpha_s(Q^2_0)/4\pi$, and the standard definition
for moments $M_{i,n} = \int_0^1 dx x^{n-2} F_i (x)$.
For more details on the notation
see ref. \cite{2} and references therein}
\begin{eqnarray}
 M_{2,n}(Q^2)   =   \sum_j M_{2,n}^j(Q_0^2)
       \left[ \frac{\a}{\ao} \right]^{d_j^n}
\left[ 1 +
(\a - \ao) \left(Q_n^{(1)j} + B_{2,n}^{(1)j}\right) \right.
\nonumber \\
+ \left. (\a^2 - \ao^2)
       \left( \tilde Q_n^{(2)j} + B_{2,n}^{(2)j} +
              Q_n^{(1)j}B_{2,n}^{(1)j} \right)
+ (\ao - \a)\ao ( Q_n^{(1)j} + B_{2,n}^{(1)j} )^2   \right] \, ,
\nonumber \\
\label{eq:3L1} \\
 M_{L,n}(Q^2)   =   \sum_j M_{L,n}^j(Q_0^2)
       \left[ \frac{\a}{\ao} \right]^{d_j^n+1}
\left[ 1 + (\a - \ao)
\left(Q_n^{(1)j} + R_{L,n}^{(2)j}\right) \right.
\nonumber \\
 +  \left. (\a^2 - \ao^2)
         \left( \tilde Q_n^{(2)j} + R_{L,n}^{(3)j} +
                 Q_n^{(1)j}R_{L,n}^{(2)j} \right)
+ (\ao - \a)\ao ( Q_n^{(1)j} + R_{L,n}^{(2)j} )^2   \right] \, ,
\nonumber
\end{eqnarray}
where $j$ extends to singlet and non-singlet
contributions.

The functions $d_j^n$, $Q_n^{(1)j}$ and $\tilde{Q}_n^{(2)j}$ are
related to the
one-loop, two-loop \cite{9} and three-loop
\cite{7} anomalous dimensions ($\gamma^{(i)n}$, $i=0,1,2$).
In the non-singlet case one has
\begin{eqnarray}
d_{NS}^n & = & \frac{ \gamma_{NS}^{(0)n} }{ 2\beta_0 } \, ,
\nonumber \\
Q_n^{(1)NS} & = &  \frac{ \gamma_{NS}^{(1)n} }{ 2\beta_0 }
               - \frac{\gamma_{NS}^{(0)n} \beta_1}{2\beta_0^2}  \, ,
\label{eq:3L2} \\
\tilde{Q}_n^{(2)NS} & = & \frac{Q_n^{(2)NS}+(Q_n^{(1)NS})^2}{2}  \, ,
\nonumber \\
Q_n^{(2)NS} & = & \frac{ \gamma_{NS}^{(2)n} }{ 2\beta_0 }
     - \frac{\beta_1}{\beta_0} Q_n^{(1)NS}
     - \frac{\beta_2\gamma_{NS}^{(0)n}}{2\beta_0^2}   \, .
\nonumber
\end{eqnarray}

The quantities $B_{k,n}^{(1)j}$ ($B_{k,n}^{(2)j}$)
and $B_{L,n}^{(2)j}$ ($B_{L,n}^{(3)j}$)
are the
NLO (NNLO)
Wilson coefficients of $F_2$ and $F_L$  respectively.
References for the two-loop coefficients can be found for example in
\cite{2}, while the new three-loop results are in
\cite{5}
and \cite{7}. In eq. (\ref{eq:3L1})
$R_{L,n}^{(i)NS} =  B_{L,n}^{(i)NS}/B_{L,n}^{(1)NS}$
for $i=2,3$.
$\beta_i$ are the coefficients
of the renormalization group beta function. For the scheme dependent
$\beta_2$ we use the $\ms$ expression given in ref. \cite{TARASOV}

In eq. (\ref{eq:3L1}) the moments $M_2(Q_0^2)$ and $M_L(Q_0^2)$ are
related at NLO by
\begin{eqnarray}
M_{L,n}^j(Q_0^2) & = & M_{2,n}^j(Q_0^2) \ao B_{L,n}^{(1)j}
\left( 1 + \ao [R_{L,n}^{(2)j} - B_{2,n}^{(1)j}] \right.
\nonumber \\
  & + & \left. \ao^2 [R_{L,n}^{(3)j} - B_{2,n}^{(2)j}
- B_{2,n}^{(1)j} ( R_{L,n}^{(2)j} - B_{2,n}^{(1)j} )] \right) \, ,
\label{eq:3L3}
\end{eqnarray}
and the three-loop coupling constant is obtained by solving
the transcendental equation
\begin{eqnarray}
\beta_0 \ln \frac{Q^2}{\Lambda^2} =
\frac{1}{\a} + \frac{\beta_1}{\beta_0} \ln (\beta_0 \a)
+ \left(\frac{\beta_2}{\beta_0} - \frac{\beta_1^2}{\beta_0^2}\right) \a
\, ,
\label{eq:3L5}
\end{eqnarray}

The only piece unknown in the perturbative calculation
is the moment of $F_2$ at some energy scale $Q_0$.
In the non-singlet case it is parametrized by
\begin{eqnarray}
M_{2,n}(Q_0^2) & = &
\int_0^1 \; dx x^{n-2 }
    \left[ A x^{\alpha} (1-x)^{\beta} (1 + \gamma x)
    \right]  \, ,
\label{eq:3L4}
\end{eqnarray}
where $A$, $\alpha$, $\beta$ and $\gamma$, in addition to
the energy scale $\Lambda$, have to be obtained from
fits to experimental data.

The expressions for the moments $M_{2,n}(Q^2)$ and $M_{L,n}(Q^2)$ in the
scheme-invariant (SI) approach can be found in \cite{2}, where the
equations for the SI coupling constants $a_{k,n}^{(i)j}$, for i=1 (NLO),
were also given.
At three loops
they obey the following
equations:
\begin{eqnarray}
\beta_0 \ln \frac{Q^2}{\Lambda^2_{\ms}} - r_{k,n}^{(1)j}=
\frac{1}{a_{k,n}^{(2)j}} +
\frac{\beta_1}{\beta_0} \ln (\beta_0 a_{k,n}^{(2)j})
+ \left(\frac{\tilde \beta_2}{\beta_0} - \frac{\beta_1^2}{\beta_0^2}\right)
a_{k,n}^{(2)j},
\label{eq:3L6}
\end{eqnarray}
where
\begin{eqnarray}
\tilde \beta_2 = \beta_2 -r_{k,n}^{(1)j}\beta_1 +
\beta_0(r_{k,n}^{(2)j} - (r_{k,n}^{(1)j})^2),
\label{eq:3L66}
\end{eqnarray}
and
\begin{eqnarray}
r_{2,n}^{(1)j}&=&\frac{Q_n^{(1)j} + B_{2,n}^{(1)j}}{d_j^n},
\nonumber \\
r_{L,n}^{(1)j}&=&\frac{Q_n^{(1)j} + R_{L,n}^{(2)j}}{d_j^n+1},
\nonumber \\
\label{eq:3L666} \\
r_{2,n}^{(2)j}&=&\frac{\tilde Q_n^{(2)j} + B_{2,n}^{(2)j} +
Q_n^{(1)j}B_{2,n}^{(1)j}}{d_j^n} -
\frac{d_j^n-1}{2} \left(r_{2,n}^{(1)j}\right)^2,
\nonumber \\
r_{L,n}^{(2)j}&=&\frac{\tilde Q_n^{(2)j} + R_{L,n}^{(3)j} +
Q_n^{(1)j}R_{L,n}^{(2)j}}{d_j^n+1} -
\frac{d_j^n}{2} \left(r_{L,n}^{(1)j}\right)^2.
\nonumber
\end{eqnarray}
In the two-loop approximation the couplings
$a_{k,n}^{(1)j}$
obey similar equations to eq. (\ref{eq:3L6})
but without the term linear in $a_{k,n}^{(1)j}$.

%
With the statistic of present experiments
it has been shown that structure
functions ($F$) can be reliably
reconstructed from their moments with the help
of orthogonal Jacobi polynomials
$\Theta_{k}^{\alpha \beta}(x)$
(see \cite{11} and references therein)
\begin{eqnarray}
F(x,Q^2) & \sim & x^{\alpha} (1-x)^{\beta}
\sum_{k=0}^{N_{max}} a_k(Q^2) \Theta_{k}^{\alpha \beta}(x) \, ,
\label{eq:H1}
\end{eqnarray}
where the coefficients $a_k$ are connected
to the moments of $F$ through
\begin{eqnarray}
a_{k}(Q^2) & = & \sum_{j=0}^{k} C_{j}^{(k)}(\alpha,\beta) M_{j+2}(Q^2)
\, .
\label{eq:H2}
\end{eqnarray}

This method has been used as a simple and fast alternative
(in computer consuming time)
to solve the evolution equations,
but here it is specially useful because
only the moments are known instead of the explicit
analytical result in $x$.

Notice that for the non-singlet anomalous dimensions
and $F_L$ Wilson coefficients only the first four even moments
have been calculated (see above)
while the polynomial reconstruction
requires also the odd ones
and at least 8 in total in order
to get an accurate result \cite{12}.
We solve this problem by obtaining the values
for n = 3, 5 and 7 from smooth interpolation (see fig. 1).
This interpolation is based on theoretical investigations
(see \cite{kaz,kot}) where it was showed that the
n-dependent functions in the Wilson coefficients
and anomalous dimensions
can be transformed to polygamma-functions
(and some others associated to it)
which are continuous in its arguments.
Thus, the analytical
extension from even values of the argument
to odd ones (see \cite{kaz})
and moreover to non integer ones (see \cite{kot}) can be done
without any problem.

The three-loops anomalous dimension can also be extrapolated to
$n>8$ if one assumes that the
behavior observed
for the first eight moments,
which is similar to the two-loop
anomalous dimension (see fig. 2), remains
at higher $n$.
In that case one can use the relation
\begin{eqnarray}
\gamma_{NS}^{(2)n} = \gamma_{NS}^{(1)n}
 \frac{ \gamma_{NS}^{(2)2} }{ \gamma_{NS}^{(1)2} }
\label{eq:EXTRAPOLATION}
\end{eqnarray}
to extend the calculation to higher $n$ (dashed line
in fig. 1).

There is not, however, an analogous relation
to extrapolate
the Wilson coefficients, of which we only know the values
at n=2,...,10 for $F_2$ and $n=2$, ..., $8$ for $F_L$. Thus
the sum in eq. (\ref{eq:H1})
was restricted to N$_{max}$=6, i.e.
moments from n=2 to n=8.

The inclusion of target mass (TM) corrections is
not possible at NNLO, because
one would need to know the moments for
$n>8$ (see eq. (23) from \cite{2}).
Also, we did not consider TM effects in the rest of
the analysis
in order to make a meaningful
comparison between the results.

We have performed a series of non-singlet fits to $F_2$
BCDMS deuterium \cite{10} and
proton \cite{10PROTON} data, and also to
SLAC reanalized $R=F_L/2xF_1$ data from deuterium \cite{9.5}.
Structure functions were calculated
at LO, NLO, NNLO and
in both scheme invariant, NLO (SI-N) and NNLO (SI-NN),
approaches for comparison\footnote{
The results at LO, NLO and SI-N for deuterium here
are different from those
in a similar analysis in reference \cite{2} due to
the inclusion of target mass corrections in \cite{2}.
We have also checked that the using
of a higher $N_{max}$ in eq. (\ref{eq:H1}), when
it is possible, does not affect
significantly the results}.
Throughout this work we have considered $n_f=4$
in the theoretical
expressions, and the reference point in eq.
(\ref{eq:3L4}) was fixed to $Q_0^2=50$ GeV$^2$.

In the analysis of $R$ we have fitted simultaneously
$F_2$ in order to fix as much as possible
the scale $\Lambda$ with the help of the more
precise BCDMS data.

The analysis has been restricted to data
with $Q^2>5$ GeV$^2$ to avoid the kinematic
region where higher twist effects could
affect $F_2$ because they are not under theoretical control.
In $R$ this cut excludes the region affected
by corrections of order $1/Q^4$ or higher.
Also we analyze the region $x>0.35$ where the influence
of gluons can be neglected and the non-singlet evolution
is justified.
In addition, it has been used $F_2$ data
with $y>0.16$, being $y$
the fraction of energy in
the laboratory system
transferred to the nucleus,
in order to exclude a region of large
systematic errors.
In practice it produces a different $Q^2$ cut
for each $x$ bin which excludes
even more the region (high $x$
and low $Q^2$) where power corrections could affect.

Table 1 summarizes the results obtained.
In the fits to only $F_2$ we find that there is
a small
difference between $\chi^2$ obtained from
different theoretical predictions, which shows
that data does not discriminate
between them.
However, in the analysis of
the more precise proton data,
the LO fit gives the worst agreement.

There are also differences in
the central value of $\Lambda$ of the fits.
The scale obtained in the SI-N analysis,
even though within the errors,
is significantly smaller than the NLO result
(12 $\%$ for deuterium and 9 $\%$ for proton).
This decreasing was already observed in earlier similar analyses
(see for example ref. \cite{2} and references therein)
and interpreted as the result of the influence
of higher order corrections which are better taken
into account by the scheme invariant
treatment\footnote{ We do not clarify the concrete SI procedure
here. This may be Stevenson's PMS \cite{14}, Grunberg's EC method
\cite{15} and other procedures (see for example \cite{2,16} and
its references), which seems to lead to similar conclusions
(see \cite{12.5,13})}.
Now, one can test this supposition by
comparing the results from the analysis NNLO and SI-N.
One see in table 1 that the value
of $\Lambda$ at NNLO is lower than at
NLO even though the decreasing is not so
pronounced as it is in the SI-N case. We think however that
it shows the trend predicted by SI-N. This behavior
has also been observed in another independent
analysis \cite{13.5}.

We have also fitted $F_2$ in the SI-NN case.
In this case
equations (\ref{eq:3L6}) above and also
eq. (16) and (17)
in \cite{2} have been used in the calculation.
The results presented in table 1 show that there is no
appreciable variation in $\Lambda$
with respect to the standard NNLO approximation.

For the combined fits to $F_2$ and $R$
(without twist corrections)
the situation is unfortunately more
complicated.
Some of the predictions are fitting better $R$
than $F_2$, giving higher $\Lambda$ values
that can not be compared.
Notice for example the poor partial fit to $F_2$
in the SI-N analysis.
However, if target mass corrections
are included in $R$, one gets a very
good fit to $F_2$ (see for example \cite{2}).
It shows that these corrections,
which are very large in $R$ \cite{2,13.6},
are essential for a meaningful combined analysis.

Comparing the results obtained from
the NLO and NNLO analyses,
one can also notice, as in the fits to only $F_2$,
a small decreasing in $\Lambda$ in the NNLO case.

In the SI-NN fits
we were not able to
converge to
a reasonable
value of $\chi^2$.
This could be due to
large $\alpha^3$ corrections in $F_L$, or again an
effect due to the absence of TM corrections.
Also it is possible that the use of $N_{max}=6$
to reconstruct $F_L$ from the
moments is not sufficient in this case.
This point will be clarify when
$\alpha^3$ correction
to $F_L$ are known at larger values of $n$.

Finally, trying to improve the agreement with $R$
data we have included
twist-4 corrections
in the calculation of $F_L$.
That contribution has been
simply parametrized by \cite{TWIST4}
\begin{eqnarray}
 M_{L,n}^{twist-4}(Q^2) = 8 \frac{\kappa^2}{Q^2} M_{2,n}(Q^2)
\, ,
\label{eq:TWIST}
\end{eqnarray}
where the energy scale $\kappa$
is the only unknown parameter
which has to be determined from data.

In all cases we found a significant improvement of
the agreement with $R$ data. The scale $\kappa$
is higher than in previous analysis
\cite{2} because in the present work it contains
part of the effect due to target mass corrections.
The value of $\Lambda$ from
the NNLO analysis is lower than the NLO result,
as in the others cases analyzed above,
and equal to the SI-NN result.
For the scale $\kappa$, this behavior is obtained
only in the proton case.

In conclusion, it has been performed for the first time a
NNLO analysis of $F_2$ and R data.
We found a slight decrease of $\Lambda$ when the NNLO
expression for the moments is used, which brings the
result closer to scheme invariant predictions.
This behavior is also in agreement
with similar analyses of
$R$-ratio in $e^+e^-$ annihilation, the $\tau$-lepton decay rate
(see \cite{12.5}) and the Gross-Llewellyn Smith sum rule in \cite{13}.

%
%
%
%

\vspace{1cm}
\hspace{1cm} \Large{} {\bf Acknowledgements}    \vspace{0.5cm}

\normalsize{}
One of the authors (G.P.)
acknowledges the financial
support from the `Comision Interministerial de Ciencia y
Tecnolog\'\i a', Spain.
Another of the author (A.V.K.) is grateful to
A.L.Kataev, S.A.Larin and T. van Ritbergen
for useful discussions. The authors are also grateful to S.A.Larin for the
information about the preprint \cite{7}.

%
%

\newpage
\section*{Figure captions}

\begin{enumerate}

\item Three-loop coefficient of the non-singlet
anomalous dimension in the $\ms$ renormalization scheme.

\item One, two and three-loop coefficients of the non-
singlet anomalous dimension in the $\ms$ renormalization scheme
normalized to the value
of the first moment $n=2$.


\end{enumerate}

\section*{Table captions}

\begin{enumerate}

\item
Results of non-singlet fits to $F_2$ proton
\cite{10} and deuterium \cite{10PROTON} data from BCDMS,
and to $R$ data from SLAC \cite{9.5}. Errors given are
statistical.

\end{enumerate}

\newpage


%
%
\begin{table}
%
%
\begin{tabular}{|c|c||c|c|c|c||} \hline
               Deuterium           &
                                    &
                  $\Lambda_{\ms}$ (MeV) &
                  $\kappa$ (MeV)      &
                  $\chi^2$($F_2$)/dof  &
                  $\chi^2$($R$)/dof     \\
\hline\hline
%
%
                    $F_2$   &
                      LO     &
             $ 182 \pm 32 $   &
                               &
                   $63.5/65$    &
                                \\
                                 &
                     NLO          &
             $ 182 \pm 30 $        &
                                    &
                   $62.7/65$         &
                                     \\
                            &
                  SI(N)      &
             $ 159 \pm 25 $   &
                               &
                   $62.4/65$     &
                                 \\
                                 &
                    NNLO          &
             $ 168 \pm 27 $        &
                                    &
                   $62.5/65$         &
                                     \\
                            &
                     SI(NN)  &
            $ 164 \pm 26 $    &
                               &
               $62.4/65$        &
                                 \\
\hline
%
%
             $F_2$ and $R$   &
                      LO      &
             $ 223 \pm 35 $    &
                                &
                   $65.1/65$       &
                   $91.5/43$        \\
                                 &
                       NLO        &
             $ 235 \pm 34 $        &
                                    &
                   $65.5/65$           &
                   $90.9/43$           \\


                            &
                      SI(N)  &
             $ 238 \pm 28 $   &
                               &
                 $70.8/65$        &
                 $66.3/43$         \\
                                 &
                      NNLO        &
             $ 218 \pm 27 $        &
                                    &
                   $65.5/65$           &
                   $79.9/43$           \\
\hline
%
%
       $F_2$ and $R$             &
                          LO      &
             $181\pm31$       &
                 $207\pm13$    &
                      $63.5/65$    &
                        $41.4/43$    \\

        (Including twist-4)      &
                   NLO            &
             $180\pm29$       &
                 $198\pm14$    &
                      $62.7/65$    &
                        $41.5/43$    \\
                            &
                      SI(N)   &
             $157\pm25$       &
                 $184\pm17$    &
                      $62.4/65$    &
                        $41.4/43$    \\
                                 &
                   NNLO           &
             $159\pm21$       &
                 $200\pm12$    &
                      $62.8/65$    &
                        $44.6/43$    \\
\hline
\hline
                  Proton            &
                                    &
                                     &
                                     &
                                      &
                                       \\
\hline\hline
%
%
                    $F_2$   &
                      LO     &
              $171 \pm 27$    &
                               &
                   $51.2/60$    &
                                \\
                                 &
                     NLO          &
             $ 175 \pm 26 $        &
                                    &
                   $47.6/60$         &
                                     \\
                            &
                      SI(N)  &
             $ 159 \pm 23 $   &
                               &
                   $46.0/60$     &
                                 \\
                                 &
                    NNLO          &
             $ 168 \pm 25 $        &
                                    &
                   $46.4/60$         &
                                     \\
                            &
                     SI(NN)  &
             $ 169 \pm 25 $   &
                               &
                   $45.8/60$    &
                                 \\
\hline
%
%
             $F_2$ and $R$   &
                      LO      &
             $ 200 \pm 30 $    &
                                &
                   $52.8/60$      &
                   $92.3/43$        \\
                                 &
                       NLO        &
             $ 222 \pm 29 $        &
                                    &
                   $50.0/60$           &
                   $82.8/43$           \\


                            &
                      SI(N)  &
             $ 231 \pm 25 $   &
                               &
                 $54.5/60$        &
                 $59.2/43$         \\
                                 &
                      NNLO        &
             $ 209 \pm 23 $        &
                                    &
                   $48.8/60$           &
                   $79.7/43$           \\
\hline
%
%
       $F_2$ and $R$             &
                          LO      &
              $ 168 \pm 27$          &
               $  205 \pm 14$         &
                $    52.0/60$          &
                      $ 43.0/43$        \\
        (Including twist-4)      &
                   NLO            &
              $ 175 \pm 26$          &
               $  195 \pm 14$         &
                   $ 48.2/60 $         &
                    $   43.5/43 $       \\
                                &
                      SI(N)       &
              $ 158 \pm 23 $         &
               $  179 \pm 17 $        &
                 $   46.6/60  $        &
                    $   42.7/43 $       \\
                                 &
                   NNLO           &
             $  158 \pm 22 $         &
               $  189 \pm 14 $        &
                $    46.5/60 $         &
                     $  47.8/43 $       \\
\hline
\end{tabular}
%
\end{table}

\end{document}